\documentclass[traditabstract]{aa} % for the abstract without structuration 
\usepackage{ulem}
\usepackage{amsmath}
\usepackage{graphicx}
\usepackage{txfonts}
\usepackage{natbib}
\usepackage{fixltx2e}

\bibpunct{(}{)}{;}{a}{}{,} 

\newcommand\asec{^{\prime\prime}}
\newcommand\amin{^{\prime}}
\newcommand\sect{\S}

\newcommand\mtx{} 

\begin{document}

\title{Redshift evolution of the 1.4 GHz volume averaged radio luminosity
  function in clusters of galaxies}
\titlerunning{Galaxy cluster radio luminosity function}

\author{M.~W.~Sommer\inst{1}\fnmsep\thanks{\email{mnord@astro.uni-bonn.de}}
  \and K. Basu\inst{1,2} \and F. Pacaud\inst{1} \and F. Bertoldi\inst{1} \and
  H. Andernach\inst{3,1}}

\institute{Argelander-Institut f\"{u}r Astronomie, Auf dem H\"ugel 71,D-53121
  Bonn, Germany \and Max Planck-Institut f\"ur Radioastronomie, Auf dem
  H\"ugel 69, D-53121 Bonn, Germany \and Permanent address: Departamento
  de Astronom\'{i}a, Universidad de Guanajuato, AP~144, Guanajuato CP 36000,
  Mexico}

\date{Received ...; accepted ...}

\abstract { By cross-correlating large samples of galaxy clusters with
  publicly available radio source catalogs, we construct the
  volume-averaged radio luminosity function (RLF) in clusters of
  galaxies, and investigate its dependence on cluster redshift and
  mass. In addition, we determine the correlation between the cluster
  mass and the radio luminosity of the brightest source within 50 kpc
  from the cluster center. We use two cluster samples: the optically
  selected maxBCG cluster catalog and a composite sample of X-ray
  selected clusters. The radio data come from the VLA NVSS and FIRST
  surveys. We use scaling relations to estimate cluster masses and
  radii to get robust estimates of cluster volumes. We determine the
  projected radial distribution of sources, for which we find no
  dependence on luminosity or cluster mass.  Background and foreground
  sources are statistically accounted for, and we account for
  confusion of radio sources by adaptively degrading the resolution of
  the radio source surveys. We determine the redshift evolution of the
  RLF under the assumption that its overall shape does not change with
  redshift. Our results are consistent with a pure luminosity
  evolution of the RLF in the range $0.1 \leq z \leq 0.3$ from the
  optical cluster sample.  The X-ray sample extends to higher redshift
  and yields results also consistent with a pure luminosity evolution.
  We find no direct evidence of a dependence of the RLF on cluster
  mass from the present data, although the data are consistent with
  the most luminous sources only being found in high-mass systems.
}
\keywords{Galaxies: clusters: general -- Radio continuum: galaxies --
  Galaxies: active -- Galaxies: evolution}

\maketitle

\section{Introduction}
\label{sec:intro}

The study of radio sources inside galaxy clusters is becoming
increasingly important in the context of large-area Sunyaev-Zel'dovich
effect \citep[SZE,][]{1972A&A....20..189S, 1999PhR...310...97B}
cluster surveys \cite[e.g.][]{2003NewAR..47..933S,
  2009ApJ...694.1610H, 2010ApJ...722.1180V, 2010arXiv1012.1610M},
especially in the case of single-frequency SZE surveys and where the
resolution is matched with the typical cluster size.  Powerful radio
sources are generally associated with early type galaxies, and as the
latter preferentially reside in clusters, it is expected that radio
sources can alter the SZE decrement in both position an depth.

Apart from contamination of SZE surveys, a study of the radio-loud
population of AGN inside galaxy clusters sheds light on the
interaction between cluster cooling flows and AGN heating/feedback
scenarios \citep[e.g.][]{2009ApJ...701...66M}.  Analyzing the radio
properties of galaxy clusters can provide direct evidence of heating
of the intra-cluster medium (ICM) by AGN through various stages of
cluster formation \citep[e.g.][]{2009ApJ...705..854H}.

Three pieces of information are crucial to assess the radio source
contamination of SZE observations: (1) the distribution of sources as
a function of projected distance from the cluster center; (2) the
brightness distribution of sources and its redshift evolution; and (3)
the spectral energy distributions (SEDs) of the sources.  A convenient
way of dealing with the first two points is to construct a radio
luminosity function (RLF), i.e. the number density of sources as a
function of luminosity, averaged over the cluster volume. This is the
objective of the present paper.

\citet{1996AJ....112....9L} first constructed the RLF for radio
galaxies in clusters using a 20 cm VLA survey of Abell clusters.
Similar studies were carried out by \citet{2004ApJ...600..695R} for
seven nearby galaxy clusters, and by \citet{2004A&A...424..409M} for a
much larger sample of 951 Abell clusters at low
redshifts. \citet{1999AJ....117.1967S} used a sample of 19 X-ray
selected clusters at $0.2<z<0.8$ to constrain the evolution of radio
galaxies, and comparing their results with those of
\citet{1996AJ....112....9L}, they found no evidence for an evolution
of the radio-loud AGN population. Similar conclusions were drawn by
\citet{2007AJ....134..897C} using a sample of massive clusters
detected through the SZE decrement at 30 GHz.  Although
\citet{2006A&A...446...97B} found indications of redshift evolution in
both slope and amplitude of the cluster RLF when comparing VLA
observations of a sample of 18 X-ray selected clusters with the
results of Stocke et al., no definitive conclusions were drawn on the
RLF redshift evolution.

Recent high-resolution X-ray imaging of clusters have provided
evidence that the AGN fraction inside clusters rapidly evolves with
redshift \citep{2007ApJ...664L...9E, 2009ApJ...701...66M}. These
results are affected by small number statistics, since the AGN
fraction in any single cluster is very small and can also vary with
other cluster properties, such as the velocity dispersion
\citep{2008ApJ...682..803S}. The volume-averaged radio luminosity
function is therefore expected to be a more robust indicator of
redshift evolution.  Correlating X-ray selected clusters with optical
and IR selected AGN, \cite{2009ApJ...694.1309G} confirmed that the AGN
excess near cluster centers increases with redshift. A parametric
modeling of this positive evolution, which can be used to predict the
radio AGN contamination in SZE cluster surveys, was not considered in
these studies.

\citet[][hereafter LM07]{2007ApJS..170...71L} used a sample of 573
X-ray selected clusters to estimate masses and virial radii, yielding
a more physically meaningful cluster volume for the RLF than obtained
when using a constant cluster radius. We adopt that method in this
paper. LM07 found a narrow radial distribution of radio sources, which
we reproduce in this work.

This study focuses on the redshift evolution of the RLF and its
decoupling from a possible dependence on cluster mass. We use publicly
available radio source catalogs at 1.4 GHz to identify radio sources
associated with clusters of galaxies in a large sample of optically
selected clusters \citep{2007ApJ...660..239K} and a composite sample
of ROSAT selected X-ray clusters.  Studies of radio point sources in
clusters at higher frequencies were carried out by, e.g.,
\cite{1998AJ....115.1388C}, \cite{2007AJ....134..897C} and
\cite{2009ApJ...694..992L}, and the extrapolation of the RLF from 1.4
GHz to frequencies relevant for SZE surveys was investigated by, e.g.,
\citet{2004A&A...424..409M} and by LM07.  These two aspects are not
pursued in this work.

This paper is organized as follows: in \sect\ref{sec:samples} we
describe the samples of radio sources and cluster of galaxies. We
discuss the estimation of cluster masses from observable properties
for clusters selected from optical and X-ray cluster catalogs, and
describe how radio luminosities are derived from flux densities. The
radial density profile of radio sources associated with clusters of
galaxies is discussed in \sect\ref{sec:profiles}.  In
\sect\ref{sec:mlcorr}, we investigate the correlation between cluster
mass and the luminosity of the brightest radio source in the central
region of the cluster. The main part of the paper is
\sect\ref{sec:lf}, where the 1.4 GHz volume-averaged radio luminosity
in galaxy clusters is derived from our sample. The method of computing
the RLF is described, and the statistical treatment of sources not
associated with the clusters is discussed. Confusion of sources is
discussed, and the redshift evolution of the RLF is derived. We
summarize our main results and offer our conclusions in
\sect\ref{sec:concl}.

Where not otherwise noted, we use the cosmological parameters from the WMAP
5-year cosmology \citep{2009ApJS..180..330K} with $h=0.705$, $\Omega_{m} h^2 =
0.136$ and $\Omega_{\Lambda}=0.726$.

%%%%%%%%%%%%%%%%%%%%%%

\section{Cluster and radio source samples}
\label{sec:samples}

In this section we describe the galaxy cluster samples and 1.4 GHz radio
source catalogs used in this paper.

\subsection{Cluster samples}
\label{sec:samples:clusters}

To demonstrate a redshift evolution of the RLF in the best possible
way, we select clusters with a wide range of redshifts from publicly
available optical and X-ray surveys.  To make robust estimates of the
volume averaged RLF, we derive cluster masses and radii using scaling
relations. The mass within a region whose mean density is 200 times
the critical density of the universe at the cluster redshift,
$M_{200}$, has been found to be a good estimator of the virial mass
\cite[e.g.][]{2001A&A...367...27W}, and is used here together with the
corresponding radius, $r_{200}$, which we use to define the typical
scale of the cluster.

\subsubsection{Description of the samples}
\label{sec:samples:clusters:descr}

Because of its large sample size, we use as our main sample the
optical maxBCG catalog \citep{2007ApJ...660..239K}, which contains a
total of 13,823 clusters extracted from the Sloan Digital Sky Survey
(SDSS). This is by far the largest homogeneous publicly available
cluster sample, and its redshift range $0.1 < z < 0.3$, corresponding
to approximately $2.5 \times 10^9$ years of cosmic time, is large
enough to allow an investigation of the redshift dependence of the
RLF.

\begin{figure}
  \centering
  \includegraphics[width=\columnwidth]{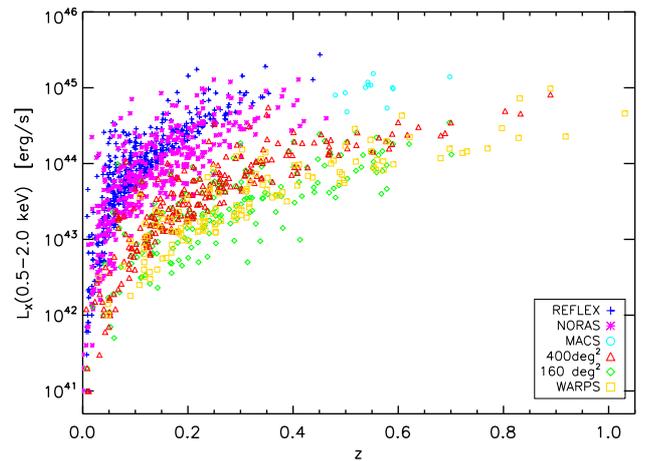}
  \caption{Distribution of X-ray selected clusters of
      galaxies in the $L_X-z$ plane.}
  \label{fig:xraysel}
\end{figure}

To allow a comparison with X-ray selected clusters, we have gathered
in one large sample most of the published cluster detections from the
ROSAT mission. Data products from this satellite remain the reference
to date since only few new massive systems have been reported from
more recent observatories.  Our source list includes the ROSAT all-sky
survey catalogs from NORAS and REFLEX \citep{2000ApJS..129..435B,
  2004A&A...425..367B}, and some serendipitous catalogs extracted from
the archives of pointed observations: the 160deg$^2$
\citep{1998ApJ...502..558V}, the 400deg$^2$
\citep{2007ApJS..172..561B} and WARPSI/II \citep{2007ApJS..172..561B,
  2008ApJS..176..374H} catalogs. We also added the complete sample of
12 high redshift very luminous systems of \cite{2007ApJ...661L..33E}.
In total this yields 1177 X-ray selected clusters. The number of
clusters from the respective catalogs and the corresponding flux
limits are given in Table~\ref{tab:xraysel}.
\begin{table*}[Ht!]
  \caption{Our composite X-ray cluster sample. The third column is the effective flux limit, in the [0.5-2] keV energy band, estimated from our recomputed source parameters. 
  }  % title of Table
\label{tab:xraysel}      % is used to refer this table in the text
\centering                          % used for centering table
\begin{tabular}{lrr}        % centered columns (4 columns)
\hline
\hline
Catalog & Number of clusters & Limiting flux ($\text{erg} \, \text{s}^{-1} \, \text{cm}^{-2}$) \\
\hline
REFLEX     &      447   &     2.1e-12     \\
NORAS       &     371   &     1.2e-12     \\
160deg2      &    221   &     8.0e-14     \\
400deg2      &    242   &     1.4e-13     \\
WARPS1+2     &    124   &     8.0e-14     \\
MACS(z$>$0.5)  &     12   &     7.0e-13     \\
\hline 
\end{tabular}
\end{table*}
Mixing these catalogs enables us to sample a large fraction of the
$L_X-z$ plane, as can be seen in Fig.~\ref{fig:xraysel}.

\subsubsection{Cluster masses and radii}
\label{sec:samples:clusters:scaling}

For the maxBCG sample, we estimate halo masses from the scaled
richness parameter $n_{\text{gal}}^{R200}$. \cite{2007ApJ...660..239K}
and \cite{2007ApJ...669..905B} find that this measure correlates well
with the galaxy velocity dispersion, and thus also with halo
mass. Indeed, \cite{2010MNRAS.404.1922A} found that optical richness
and X-ray luminosity perform similarly well in predicting cluster
masses. The relation between $n_{\text{gal}}^{R200}$ and $M_{200}$ has
been extensively studied, both using weak lensing
\citep{2007arXiv0709.1159J,2008JCAP...08..006M,2009ApJ...703.2217S,
  2009ApJ...699..768R} and comparing to X-ray luminosities
\citep{2008MNRAS.387L..28R, 2008ApJ...675.1106R, 2009ApJ...703..601R}.

In this paper, we use the X-ray luminosity$-$optical richness relation
($L_X - n_{\text{gal}}^{R200}$) found by \cite{2008ApJ...675.1106R} to
derive the expected X-ray luminosities as
\begin{equation}
\label{eq:rykoff}
L_{x}(R_{200}) = e^{\alpha} \left(
  \frac{n_{\text{gal}}^{R200}}{40} \right)^{\beta} \left( \frac{1+z}{1.23} \right)^{\gamma} \times 10^{42} \, h^{-2} 
\,\, \text{ergs} \,\, \text{s}^{-1},
\end{equation}
where $\alpha = 3.90 \pm 0.04$, $\beta = 1.85 \pm 0.05$ and $\gamma =
6.0 \pm 0.8$.  We scale the derived X-ray luminosities to mass using a
self-similar redshift evolution, in the same way as described below
for the X-ray sample.

For the X-ray sample, the heterogeneity of the data makes it difficult
to straightforwardly estimate the cluster masses in a uniform way. The
most basic observable, available for all the sub-samples although in
different bands ($[b_{min},b_{max}]$), is the total cluster X-ray
flux.  We thus apply the scaling relation of
\cite{1999MNRAS.305..631A}, linking the average gas temperature $T$ to
the bolometric X-ray luminosity $L_{\text{bol}}$, with a self-similar
redshift evolution to define a unique mapping between
($z$,$F_X$,$b_{min}$,$b_{max}$) and ($L_{bol}$,$T$).  For this, we
make use of the APEC spectral model \citep{2001ApJ...556L..91S} with
heavy element abundances set to 0.3 times the solar values.  The
derived cluster temperatures are translated into $M_{200}$ following
the scaling law of \cite{2005A&A...441..893A}, again assuming
self-similar evolution.

From simple geometric considerations and the definition of over-density with
respect to the critical density, it follows that $r_{200}$ is related to
$M_{200}$ by
\begin{equation}
\label{eq:m200r200}
M_{200} \equiv (4\pi/3) r_{200}^3 \, 200 \, \rho_c(z), 
\end{equation}
where $\rho_c(z)$ is the critical density of the universe at redshift
$z$.  $r_{200}$ is typically in the range $1-3$ Mpc for the clusters
in our samples.

\subsubsection{Cluster sample selection criteria}
\label{sec:samples:clusters:cuts}

In order to limit our survey to massive systems, we exclude all
clusters less massive than $5 \times 10^{13} M_{\odot}$ from our
sample. We also exclude low-redshift systems with $z<0.1$, in order to
have the same redshift cutoff in both samples. In addition, we define
a low-redshift ($0.05 < z < 0.12 $) sub-sample of the X-ray sample,
with the same mass cut (see Table \ref{tab:ccuts}), to investigate the
mass dependence of the RLF (\sect\ref{sec:lf:model:mdep}). The lower
redshift limit of the sub-sample is chosen so as to avoid excessive
overlap of low-redshift cluster fields. The upper redshift limit is
chosen as high as possible without including redshifts where the mass
limit of the flux-limited X-ray cluster sample is greater than our
mass limit of $5 \times 10^{13} M_{\odot}$, in order to avoid a bias
in the the determination of a possible mass dependence in the RLF. The
fact that the low-redshift sub-sample has a slight overlap with the
high-redshift X-ray sample is of little consequence as the two samples
are never compared directly with one another.

Not all cluster fields in our samples are covered by the NVSS and
FIRST radio surveys that are used for this study. Apart from taking
this into account, we note that due to dynamic range limitations of
the radio interferometric data, in the FIRST and NVSS catalogs some
regions on the sky around strong radio sources are plagued by
abnormally high or low source counts, often with catalog entries not
corresponding to real objects.\footnote{Typical examples are the
  fields centered on the quasars 3C273 and 3C295} We identify these
regions by counting the number of sources in a circular region with
radius $1^{\circ}$ around each cluster center, and excluding fields
where the source counts exceed or fall below the average counts in
cluster fields by more than three standard deviations.\footnote{The
  $1^{\circ}$ radius corresponds to approximately ten times $r_{200}$
  at the median redshift and mass of the maxBCG sample, and is chosen
  so as to include all cluster sources and to have a sufficient number
  of background and foreground sources to allow for robust statistics}
Note that using the average counts of FIRST and NVSS would have
underestimates the expected counts in cluster fields, where a local
surface over-density is expected. Counting sources within one degree
of maxBCG clusters (cf. \sect\ref{sec:profiles}), we find the average
density of FIRST sources in such a region to be 9\% higher than the
catalog average of 90 sources per square degree.  In the case of NVSS,
only about 4\% of the cluster fields are affected by this cut
(determined from the difference of rows 4 and 5 of
Table~\ref{tab:ccuts} in the maxBCG case).

We also exclude pairs \mtx{(or multiplets)} of clusters
where the sum of the radii ($r_{200}$ projected on the sky plane) is
greater than the angular separation between the cluster centers
\mtx{(row 2 of Table~\ref{tab:ccuts})}.

Table \ref{tab:ccuts} summarizes the selection criteria applied to the
cluster sample, and lists how many clusters in each of the final
samples are covered by the FIRST and NVSS catalogs.

\begin{table*}[Ht!]
\begin{minipage}[Ht]{\textwidth}
  \caption{\mtx{Cluster sample selection criteria}. The
    selection is cumulative in the sense that in rows 3-4, each row
    assumes all the conditions of the previous rows. Rows 5 and 6
    assume all the conditions in rows 2-4. The low-$z$ X-ray sample is
    used only to constrain the mass dependence of the RLF
    (\sect\ref{sec:lf:model:mdep}). Where not explicitly stated
    otherwise, the high-redshift samples are used in this
    paper.}  % title of Table
\label{tab:ccuts}      % is used to refer this table in the text
\centering                          % used for centering table
\renewcommand{\footnoterule}{}
\begin{tabular}{c|r|c|c|c|}        % centered columns (4 columns)
      \cline{2-5}         % inserts double horizontal lines
 & \textbf{Main sample} & \textbf{maxBCG} & \multicolumn{2}{|c|}{\textbf{X-ray}} \\    % table heading
  \cline{2-5} \cline{2-5}
 \textsl{1} & clusters in main sample & 13823 & \multicolumn{2}{|c|}{1177} \\
 \textsl{2} & clusters with sufficient separation & 12846 & \multicolumn{2}{|c|}{1121}  \\
  \cline{2-5}
 & \textbf{Sub-sample}      & & \textbf{high-$z$} & \textbf{low-$z$} \\
 & Redshift range & $0.1 \leq z \leq 0.3$ & $0.1 \leq z \leq 1.26$ & $0.05 < z < 0.12 $ \\
     \cline{2-5}                    % inserts single horizontal line
 \textsl{3} & Clusters within redshift range & 12846 & 690 & 292 \\
 \textsl{4} & Clusters with $M > 5 \times 10^{13} M_{\odot}$ & 12522 & 674 & 275 \\
  \cline{2-5}
 \textsl{5} & clusters with NVSS coverage\footnote{Excluding regions with abnormally high or abnormally low source
    counts (see text)} & 12475 & 596 & 218 \\
 \textsl{6} & clusters with FIRST coverage$^{a}$ & 11812 & 273 & 75 \\
      \cline{2-5}                              %inserts single line
\end{tabular}
\end{minipage}
\end{table*}

\subsection{Radio source samples}
\label{sec:samples:radio}

\subsubsection{Description of the samples}
\label{sec:samples:radio:descr}

To search for radio sources associated with clusters, we use the FIRST
\citep{1995ApJ...450..559B, 1997ApJ...475..479W} and NVSS
\citep{1998AJ....115.1693C} 1.4 GHz radio continuum surveys. The most
relevant properties of the NVSS and FIRST surveys for the present work
are summarized in Table \ref{tab:rcata}. We make direct use of
publicly available source catalogs from each survey. While the NVSS
catalog has the advantage of greater sky coverage
(cf. Table~\ref{tab:rcata}), thereby covering more of our cluster
fields, it has the disadvantage of a poorer resolution, resulting in
increased source confusion.

\begin{table}[Ht!]
\begin{minipage}[Ht]{\columnwidth}
\caption{Properties of the FIRST and NVSS radio continuum surveys}             % title of Table
\label{tab:rcata}      % is used to refer this table in the text
\centering                          % used for centering table
\renewcommand{\footnoterule}{}  % to avoid a line before footnotes
\begin{tabular}{r c c}        % centered columns (4 columns)
  \hline\hline                 % inserts double horizontal lines
  & FIRST & NVSS \\    % table heading 
  \hline                        % inserts single horizontal line
  effective resolution & 5$^{\prime\prime}$ & 45$^{\prime\prime}$ \\
  completeness limit & 1 mJy \footnote{catalog detection threshold} & $\sim$2.5 mJy \\
  positional uncertainty\footnote{for the brightest sources} & $ < 0.5 ^{\prime\prime}$ & $ < 1 ^{\prime\prime}$\\
  positional uncertainty\footnote{at the completeness limit/detection threshold} &
  ~1$^{\prime\prime}$  & $\sim$7$^{\prime\prime}$ \\
  total number of sources & 816,331\footnote{as of July 16, 2008} & 1,810,672\footnote{as of February, 2004} \\
  sources per square degree & $\sim$90 & $\sim$\mtx{53} \\
  area covered (deg$^2$)  & 9055  &  \mtx{33885} \\

\hline                                   %inserts single line
\end{tabular}
\end{minipage}
\end{table}

We search for FIRST and NVSS sources within a projected radius corresponding
to $3 r_{200}$ around each cluster center, with $r_{200}$ computed from
Eq.~(\ref{eq:m200r200}). In order to facilitate a robust comparison of
NVSS and FIRST sources, we exclude all sources fainter than 5 mJy, which is
well above the completeness limit of both surveys.

\mtx{In principle the use of the FIRST catalog alone would be enough
  for the purpose of this paper. We make use of the NVSS catalog at
  all stages of the analysis because it serves as a test that our
  results obtained with FIRST are robust. In particular, the NVSS data
  are important for testing our method of adaptively accounting for
  confusion, as described in \sect\ref{sec:lf:confusion}.  Because
  the two surveys have significant overlap, we do not need to ``mix''
  them in the sense of combining source counts from the two catalogs
  for any particular sample.}

\subsubsection{Radio source luminosities}
\label{sec:samples:radio:lum}

To compute radio luminosities, we assume isotropic emission, which is
not correct for any individual galaxy, but is well justified when
averaging over the entire sample.  The luminosity (power) of a radio
source is
\begin{equation}
L_{\mathrm{1.4 \, GHz}} \ = \ (4\pi \ D_L^2) \ S_{\mathrm{1.4 \, GHz}} 
\ \frac{ {\cal K}(z) } { (1+z) },
%L_{\mathrm{1.4 \, GHz}} \ = \ (4\pi \ D_L^2) \ S_{\mathrm{1.4 \, GHz}} 
%\ {\kappa}(z) \ / \ (1+z)
\label{eq:lumeq}
\end{equation} 
where $S_{\mathrm{1.4 \, GHz}}$ is the angular integrated flux density
taken from the VLA catalog (FIRST or NVSS), $D_L$ is the cosmological
luminosity distance and ${\cal K}(z)$ is the $k$-correction.

\mtx{The radio sources are modeled with continuum spectra of
  the form $S_{\nu} \propto \nu^{-\alpha}$, where $\alpha$ is the
  spectral slope. When computing the luminosity from flux, we have to
  account for the fact that due to the redshift, the observed flux
  corresponds to a rest frequency higher than 1.4 GHz.  For the
  computation of this so-called $k$-correction, ${\cal K}(z) $, we
  assign a spectral slope of $\alpha=0.72$ to all sources, as
  determined by \cite{2007AJ....134..897C} in the range $1.4-30$
  GHz. The effect of this correction is about $10\%$ for sources at
  $z=0.2$ and $30\%$ at $z=0.6$.}

When binning the data by luminosity to compute the RLF, we determine a
redshift limit, $z_{\rm{cut}}$, for each luminosity bin to avoid
counting cluster fields where the bin luminosity corresponds to a flux
below our chosen threshold of 5 mJy. $z_{\rm{cut}}$ is determined from
Eq.~(\ref{eq:lumeq}) and only affects the lowest luminosities
considered in the analysis.

%%%%%%%%%%%%%%%%%%%%%%

\section{Radial density distribution of radio sources}
\label{sec:profiles}

As we have no redshift information for individual radio continuum
sources, the over-density of sources toward clusters must be
quantified statistically.  We construct the stacked radial profile of
radio sources around cluster centers. Following LM07, we take
$r/r_{200}$ as the radial coordinate. The resulting radial profile
includes both cluster galaxies and the field (background and
foreground) population.  Although it cannot be decided whether
individual sources are cluster members, cluster and field sources can
be separated statistically, since the field population has a radially
constant contribution to the radial profile. \mtx{Besides determining
  the stacked radial density distribution of all radio sources, we
  also investigate whether the distribution depends on source
  luminosity and host galaxy cluster mass.}

The radial profiles are used when constructing the luminosity function by
de-projection into volume number density as described in
\sect\ref{sec:lf:method:deproj}.

\subsection{Radial model}
\label{sec:profiles:model}

\mtx{The angular offset of radio sources with respect to the cluster
  center (defined as the position of the brightest cluster galaxy
  (BCG) in the maxBCG sample) is translated into projected physical
  distance at the cluster redshift.} The physical radial distances are
stacked for all cluster fields and binned by radius.

We fit the radial profile of radio sources in each cluster sample by a
parametric model.  In general, all our radial profiles are well fit by
a model of the form
\begin{equation}
  \psi(\xi) = \psi_{\beta} \left(1+\frac{\xi^2}{\zeta^2}\right)^{-\frac{3}{2}\beta + \frac{1}{2}} \, + \,
  \frac{1}{\xi}\psi_{G} e^{-(\frac{\xi}{2 \, \sigma})^2}   \, + \,
  \psi_f,
\label{eq:proffitfun}
\end{equation}
where $\psi(\xi)$ is the projected surface density and $\xi =
r/r_{200}$. The first term of (\ref{eq:proffitfun}) has the form of an
isothermal $\beta$-model \citep{1978A&A....70..677C}, the second term
is a Gaussian, corrected for the area in an annulus, and the third
term, $\psi_f$, is the constant field density.

The isothermal $\beta$-model is described by three parameters: the
(peak) normalization $\psi_{\beta}$, the scale radius $\zeta$, and the
power law index $\beta$. \mtx{The $\beta$-model is chosen because it
  provides a good fit to our radially binned data outside $r/r_{200}
  \simeq 0.006$.}

In our fractional radial coordinates, the constant field density,
$\psi_f$, is a function of the redshift distribution of the galaxy
cluster sample. It is not physically meaningful as we have
(incorrectly) placed all radio sources at the cluster redshifts, but
as we are not interested in the field population this component may
safely be ignored.

The second term in Eq.~(\ref{eq:proffitfun}) accounts for an
additional peaked feature in the radial distribution of FIRST sources
in the maxBCG galaxy cluster sample (Fig.~\ref{fig:prof:maxbcg}). It
is well fit by a Gaussian with normalization $\psi_{G}$ and variance
$\sigma^2$. The origin of this feature is the fact that the cluster
center is taken as the position of the BCG. As discussed further in
\sect\ref{sec:profiles:res}, the underlying distribution \mtx{derives
  from the extended nature of the large majority of FIRST radio
  sources associated with the BCG.}  Note that the central Gaussian
component of $\psi(\xi)$ is considered only for the radial
distribution of FIRST sources in the maxBCG sample; in all other cases
we set $\psi_G = 0$.

For the purpose of constructing the RLF, the exact form of the fitting
function is not important; although the density profile is used in
constructing the RLF (as discussed in \sect\ref{sec:lf:method:deproj}), the
RLF is only very weakly sensitive to the parameterization of the source
density fitting function.

\subsection{Dependence on luminosity and cluster mass}
\label{sec:profiles:lmdep}

\mtx{Because the radial distribution of sources plays an important
  role in determining cluster member radio source counts within
  $r_{200}$, it is necessary to investigate whether bright radio
  sources have a radial distribution different from the distribution
  of faint sources. We compute the luminosities of FIRST sources in
  maxBCG clusters according to Eq.~(\ref{eq:lumeq}) by placing all
  radio sources at the cluster redshifts and binning the sample in
  luminosity. We determine the radial source distribution in each
  luminosity bin using the fitting function given by
  Eq.~(\ref{eq:proffitfun}).  Similarly, we investigate a possible
  dependence on cluster mass by binning the cluster sample in mass.}
  %%(\sect\ref{sec:profiles:res:lmdep}).

\mtx{Because the sample has been divided and uncertainties on individual
  profiles are greater, we fix the power law parameter $\beta$ in
  Eq.~(\ref{eq:proffitfun}) to the best-fit value $\beta=0.987$ from the
  total fit (Table~\ref{tab:profres}) in order to get reliable estimates on
  the scale radius $\zeta$.}

\subsection{Results}
\label{sec:profiles:res}

\subsubsection{Ensemble properties}
\label{sec:profiles:res:ensamble}

\begin{figure}
  \centering
  \includegraphics[width=\columnwidth]{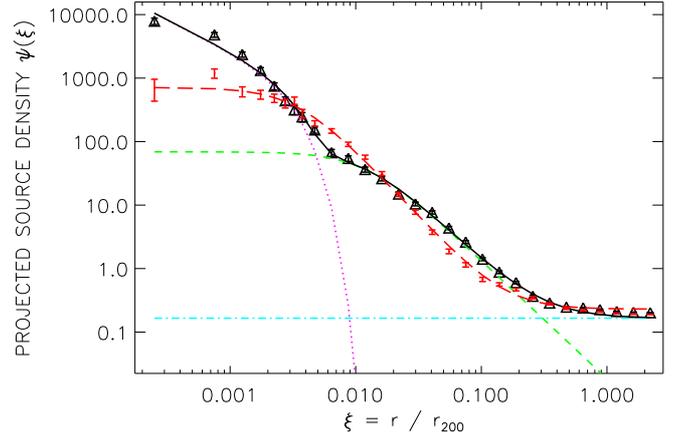}
  \caption{Radial distribution of 1.4 GHz radio sources brighter than 5
    mJy in the maxBCG cluster sample. The normalization on the $y-$axis is
    arbitrary. Radio sources selected from the FIRST catalog (triangles with
    error bars) clearly indicate three components: a narrow central peak,
    fitted with a Gaussian (dotted line); a broader distribution of sources,
    fitted with a $\beta$-profile (short-dashed line) and a
    background/foreground component (dash-dotted line). The sum of the
    components is indicated by the solid line. The radial profile derived from
    the NVSS catalog is indicated by the error bars without symbols. The fit
    to these data (red long-dashed line) does not include a Gaussian
    component.}
  \label{fig:prof:maxbcg}
\end{figure}

Figure~\ref{fig:prof:maxbcg} shows the source density profiles in
maxBCG clusters, fitted with the parametric form (\ref{eq:proffitfun})
and divided into its components.
Table \ref{tab:profres} indicates the most important parameters in the
radial fits of radio sources associated with clusters.
\begin{table*}
  \caption{Results of $\beta$-model fits to the radial distribution of sources
    associated with clusters of galaxies. The data are fitted using Eq.~(\ref{eq:proffitfun}). }
\label{tab:profres}      
\centering                          
\begin{tabular}{l r r@{}l r@{}l r@{}l r@{}l}
\hline\hline                 
Cluster sample & Source catalog &    $\beta$&      &      $\zeta$&        & $\psi_G$&     &     $\sigma$&        \\ 
\hline
maxBCG         & FIRST          & 0.987$\pm$&0.018 & 0.0115 $\pm$&0.0012  &  3.23$\pm$&0.18 & 0.00274$\pm$&0.00013 \\ 
maxBCG         & NVSS           & 1.018$\pm$&0.010 & 0.00257$\pm$&0.00019 &        $-$&     &          $-$&        \\
X-ray          & FIRST          & 0.899$\pm$&0.136 & 0.0191 $\pm$&0.0099  &        $-$&     &          $-$&        \\  
X-ray          & NVSS           & 0.911$\pm$&0.083 & 0.0198 $\pm$&0.0063  &        $-$&     &          $-$&        \\
\hline                                   
\end{tabular}
\end{table*}

\mtx{Note that the parameters ($\psi_G,\sigma$) of the
  Gaussian were fit only for the FIRST sources associated with maxBCG
  clusters, as this is the only combination of cluster and source
  catalog that yields an additional centrally peaked feature. The
  width of the peaked distribution ($\sim 3 \times 10^{-3} ~r_{200}$)
  corresponds to 0.9$\asec$ at the median distance of the maxBCG
  clusters. Although this is comparable to the maximum positional
  uncertainty of FIRST point sources ($\sim1\asec$) at the
  completeness limit, the FIRST positions are typically constrained to
  a much better accuracy, in particular at higher flux levels. Thus,
  positional uncertainties in the FIRST survey alone cannot explain
  the width of the peaked distribution.}
  
\mtx{ A visual inspection of a large fraction of $\sim$1800 FIRST
  sources with positional matches out to 15$\arcsec$ of the maxBCG
  positions revealed that the very close matches ($<~0.3\asec$) are
  dominated by compact radio sources, while the matches in the range
  $~3\asec$--$10\asec$ can be classified into three broad categories:
  (a) double sources with more or less symmetrical lobes straddling
  the BCG position, (b) triple sources with weak cores (centered on
  the BCG) and strong lobes, for which our threshold of 5~mJy picked
  up only the lobes, and (c) complex sources, usually of FRII or
  wide-angle tail type, some of which have truly spectacular radio
  morphologies and sizes up to 1$\amin$ or more. From this inspection
  we estimate that out to 10$\asec$ separation ($\sim 0.3 r_{200}$ at
  the median redshift) at least 98\% of our matches correspond to the
  BCGs (with at most ~10\% of the sources being duplicates, i.e.\
  different components or lobes of the same complex source). While for
  separations between 10 and 15$\arcsec$ (i.e.\ up to ~54 kpc at the
  median redshift of the clusters of z=0.227) the fraction of
  duplicates exceeds 30\%, our matches can be safely related to the
  BCGs in at least 90\% of the cases. We conclude that the peaked
  component in the radial distribution originates from mismatches in
  SDSS and FIRST positions caused by extended radio emission.}

Because the BCG component can be separated from the general
distribution of FIRST sources in the maxBCG sample, the dimensionless
core radius, $\zeta$, for the remaining sources is much larger than
for the NVSS sources, where the peaked central distribution is
``hidden'' in the $\beta$-model component. The distribution of sources
in the X-ray cluster sample is less peaked (larger $\zeta$) because
the cluster center definition is based on the gas distribution and not
the position of the BCG.

We note that our results pertaining to FIRST sources in maxBCG
clusters verify the radial distribution derived by
\cite{2007ApJ...667L..13C} for the same sample. As indicated in
Fig.~\ref{fig:prof:maxbcg}, our narrow binning reveals the details of
the centrally peaked component in detail and confirm its origin as the
BCG population. \mtx{The relatively high redshifts in the
  optical sample ($z \geq 0.1$) also imply that the population of star
  forming galaxies in clusters -- with a broader distribution than
  radio-loud AGN -- has been excluded by our flux limit of 5 mJy. As a
  consequence our $\beta$-model fit to the radial profile is much
  narrower compared to those obtained for the radio source
  distribution in local clusters \citep[e.g.][]{2004ApJ...600..695R,
    2004A&A...424..409M}. Similar narrow radial profiles consisting
  mostly of radio-loud AGN were also found by LM07.}

\subsubsection{Luminosity and cluster mass dependence}
\label{sec:profiles:res:lmdep}

\mtx{From their X-ray sample of galaxy clusters, LM07 found that more
  luminous radio sources are more centrally concentrated in galaxy
  clusters. As indicated in Table \ref{tab:profres:lbins}, we are
  unable to reproduce this result with our optical maxBCG sample;
  although the highest luminosity bin has a slightly smaller scale
  radius $\zeta$ of the outer profile, the result is not
  significant. Similarly, there is no statistically significant
  indication of a mass dependence of the radial density from the
  maxBCG sample of galaxy clusters (Table \ref{tab:profres:mbins}). We
  verify that the same is the case when using the NVSS sample.  }

\begin{table}
  \caption{Results of $\beta$-model fits to the radial distribution of FIRST 
    sources 
    associated with maxBCG clusters of galaxies, with the radio sources binned
    by luminosity $L$. Only the parameters $\zeta$ and $\sigma$, pertaining to
    the width of the profiles, are listed. $\beta$ has been fixed to the 
    best-fit value
    0.987 from the total fit. The data are fitted using 
    Eq.~(\ref{eq:proffitfun}).}
\label{tab:profres:lbins}      
\centering                          
\begin{tabular}{r@{}l r@{}l r@{}l}
\hline\hline                 
Luminosity& ~range                                &  $\zeta$&               & $\sigma$& \\
\hline
$\log(L \,[\text{W/Hz}])$& ~$<$ 23.8              &  0.0116 $\pm$& 0.0018 & 0.00249 $\pm$& 0.00067 \\
23.8 $\leq$ $\log(L \,[\text{W/Hz}])$& ~$<$ 24.2  &  0.0138 $\pm$& 0.0023 & 0.00319 $\pm$& 0.00054 \\
24.2 $\leq$ $\log(L \,[\text{W/Hz}])$& ~$<$ 24.6  &  0.0138 $\pm$& 0.0024 & 0.00343 $\pm$& 0.00073 \\
$\log(L \,[\text{W/Hz}])$& ~$ \geq $ 24.6              &  0.0097 $\pm$& 0.0018 & 0.00303 $\pm$& 0.00088 \\
\hline                                   
\end{tabular}
\end{table}

\begin{table}
\caption{As table~\ref{tab:profres:lbins}, but with the sample divided into
  mass bins.}
\label{tab:profres:mbins}      
\centering                          
\begin{tabular}{r@{}l r@{}l r@{}l}
\hline\hline                 
Mass& ~range                                &  $\zeta$&               & $\sigma$& \\
\hline
$\log(M/M_{\odot})$& ~$<$ 13.90             &  0.0132 $\pm$& 0.0016 & 0.00334 $\pm$& 0.00060 \\
13.90 $\leq$ $\log(M/M_{\odot})$& ~$<$ 14.15  &  0.0139 $\pm$& 0.0017 & 0.00344 $\pm$& 0.00075 \\
14.15 $\leq$ $\log(M/M_{\odot})$& ~$<$ 14.40  &  0.0125 $\pm$& 0.0011 & 0.00260 $\pm$& 0.00053 \\
$\log(M/M_{\odot})$& ~$ \geq $ 14.40              &  0.0109 $\pm$& 0.0011 & 0.00209 $\pm$& 0.00049 \\
\hline                                   
\end{tabular}
\end{table}

%%%%%%%%%%%%%%%%%%%%%%

\section{Mass-luminosity correlation}
\label{sec:mlcorr}

In this section we investigate the correlation between cluster mass and the
radio luminosity of the brightest radio source in the central region of the
cluster. \mtx{We compare the results with similar studies done with the
  optical luminosities of the brightest cluster galaxies, and discuss the
  implications for our goal of comparing the volume averaged radio luminosity
  function in clusters at different redshifts.}

\subsection{Background}
\label{sec:mlcorr:back}

\mtx{Brightest cluster galaxies (BCGs) tend to lie at the center of
  the mass distribution and are more luminous than average
  ellipticals.  Moreover, their properties correlate strongly with
  their host clusters, as seen from numerical simulations
  \citep[e.g][]{2007MNRAS.375....2D} and optical/near-IR observations
  \citep[e.g.][]{2004ApJ...617..879L,2008MNRAS.385L.103B}. It has been
  shown that BCGs are about an order of magnitude more likely than
  other ellipticals to host radio-loud AGN \citep{1991MNRAS.250..103E,
    2004ApJ...617..879L, 2007MNRAS.379..894B}.}  \mtx{Based on this,
  it is likely that the most radio-luminous galaxies in clusters are
  associated with BCGs, and in addition to the correlation between BCG
  luminosity and cluster mass one can expect that 1.4 GHz radio
  luminosities of the central radio-loud AGN are also correlated with
  cluster mass.}

\subsection{Method}
\label{sec:mlcorr:method}

We investigate the correlation of BCG radio luminosity with cluster
mass by taking the luminosity of the single brightest source within a
radius of 50 kpc from the cluster center. We reiterate that the latter
is actually the BCG position in the maxBCG sample. In the X-ray
sample, the search radius allows for offsets between the X-ray center
and the BCG position; \cite{2007ApJ...660..239K} found typical offsets
of $~50 \, h^{-1}$ kpc by cross-correlating the maxBCG catalog with
NORAS and REFLEX data.

\mtx{ Note that this selection method leads to a different minimum flux
  level for different redshift sub-samples, since we are disregarding clusters
  that do not contain any radio source above a certain flux density. This
  leads to a higher mean luminosity of the central radio source for clusters
  at higher redshift, compared to clusters at lower redshifts within the same
  mass bin (derived from the X-ray or optical mass observables). The actual
  redshift evolution of the radio luminosities of the central BCGs is hidden
  within this luminosity bias.  Considering the large scatter in the
  luminosity of the central radio source, we do not attempt to model the
  redshift evolution from this method.  }

\subsection{Results}
\label{sec:mlcorr:res}

\begin{figure}[ht!]
  \centering \includegraphics[width=\columnwidth]{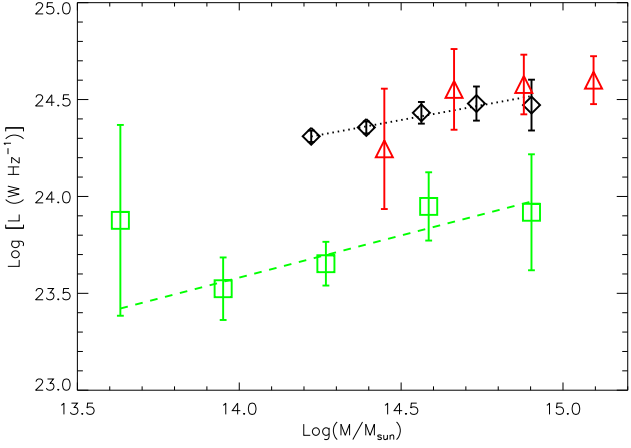}
  \caption{Correlation between radio luminosity of the brightest FIRST
    source within a projected physical radius of 50 kpc from the
    cluster center (BCG position in the maxBCG sample) and cluster
    mass for three sub-samples of galaxy clusters. Diamonds indicate
    maxBCG clusters in the redshift range $0.1 < z \leq 0.2$, and the
    best-fit correlation is indicated by the dotted
    line. 
    This is consistent with the X-ray sample in the same redshift
    range (triangles).  Fixing the slope, the correlation is
    consistent with a local sub-sample of X-ray clusters ($0.01 < z
    \leq 0.1$, squares), where the luminosity level is lower due to a
    different luminosity cutoff in this sub-sample.}
\label{fig:mlcorr}
\end{figure}

Figure \ref{fig:mlcorr} shows the correlation of luminosities of FIRST sources
with cluster mass for both our cluster samples.
Modeling the correlation with a power-law of the form $L \sim
M^{\alpha}$, the best-fit correlation from the maxBCG sample in the
redshift range $0.1 < z \leq 0.2$ is $\alpha = 0.31 \pm
0.12$. Although the X-ray sub-sample in the same redshift range is
essentially too small to constrain the power law, the data are
consistent with the maxBCG result, with $\alpha = 0.34 \pm 0.41$.  The
low-redshift X-ray sub-sample yields $\alpha = 0.44 \pm 0.27$, again
consistent with the maxBCG relation.

\mtx{The found correlation between radio luminosity and cluster mass
  is consistent with results from optical/near-IR observations of the
  BCGs.  \cite{2004ApJ...617..879L} compared the K-band near-IR
  luminosities for a sample of X-ray selected clusters and found
  $L_{\mathrm{BCG}} \propto M_{cl}^{0.26\pm
    0.04}$. \cite{2010ApJ...713.1037H} reported a shallower
  correlation from an X-ray selected sample; $L_{\mathrm{BCG}} \propto
  M_{cl}^{0.18\pm 0.07}$. Thus, luminosities in other bands show
  correlations with mass that are similar to our results.  }

\mtx{For the optical cluster sample, the correlation of radio
  luminosity and cluster mass strongly points towards increased radio
  emission from the central AGN in the BCGs of more massive clusters.
  \cite{2007ApJ...667L..13C} also computed the rest-frame 1.4 GHz
  luminosities of the BCGs by cross-correlating the maxBCG catalog
  with the FIRST survey, and noted the trend of increasing radio
  luminosity with optical luminosity of the BCGs in the $r$-band. We
  do not expand on this result further, as the radio luminosity of the
  central brightest radio source is a poor indicator of the total
  radio luminosity in clusters. \mtx{A cluster can have multiple BCGs
    as a result of its merging history,}\footnote{\mtx{After a cluster
      merger, it will take considerable time before the two initial
      BCGs cease to dominate their subgroups and one of them comes to
      dominate the merged cluster}} \mtx{and the radial source density
    profile in Fig.~\ref{fig:prof:maxbcg} suggests that the radio
    source population extends beyond the cluster core region. For
    these reasons, we now turn to the construction and modeling of the
    volume averaged radio luminosity function in galaxy clusters.}

%%%%%%%%%%%%%%%%%%%%%%

\section{The radio luminosity function}
\label{sec:lf}

\mtx{ In this section, the radio luminosity function (RLF) is derived
  for each of our samples. We start by discussing how accurate source
  counts in the cluster volume are obtained
  (\sect\ref{sec:lf:method}). In \sect\ref{sec:lf:confusion} we
  describe how source confusion is accounted for. We describe a
  parametric model for the RLF in \sect\ref{sec:lf:model}, and also
  discuss how a redshift evolution is modeled. We give the results in
  \sect\ref{sec:lf:res}.  }

\subsection{Method}
\label{sec:lf:method}

The luminosity function $\phi(L)$ is defined as the average number of
radio sources per unit physical volume of a cluster and per
logarithmic luminosity bin (``magnitude'') . We use the estimated
values of $r_{200}$ from \sect\ref{sec:samples:clusters} to define
cluster volumes, expressed in $\rm{Mpc}^3$.

\subsubsection{Source counts}
\label{sec:lf:method:counts}

Figure \ref{fig:lf:cartoon}
\begin{figure}[ht!]
\vspace*{5mm}
\includegraphics[width=\columnwidth]{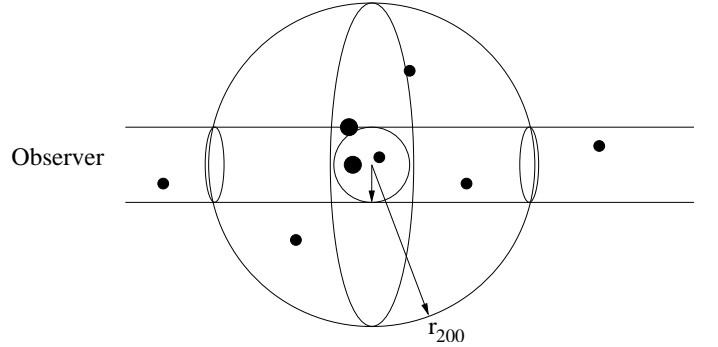}
\caption{Illustration of the
  method used to obtain source counts and volumes for the computation of the
  luminosity function of radio sources inside clusters. For each cluster the
  sampled volume is the ``line-of-sight cylinder''.  The projected number
  density in this volume is converted to volume density of sources inside the
  cluster radius ($r_{200}$) by applying a model for the volume density of
  sources.}
\label{fig:lf:cartoon}
\end{figure}
illustrates schematically how source counts are obtained. For each
cluster, a region within a projected physical radius $r = \eta \,
r_{200}$, from the center is searched for radio sources in a given
luminosity bin, using the cluster redshift to convert from flux to
luminosity.  As explained below, the search radius does not have a
systematic effect on the RLF; however, it can be chosen such as to
minimize the uncertainties of the latter. The number of sources found
within the chosen radius is $N_{\eta}^{\rm{count}}$. We need to
convert this number into a de-projected number of sources within
$r_{200}$.  For this, we need to know (i) the number of
foreground/background sources, and (ii) the radial source density
profile. In the following we discuss these issues in turn.

\subsubsection{Field subtraction}
\label{sec:lf:method:field}

To correctly model the field counts, we first need to understand the mean
surface number density of radio sources in the sky as a function of flux. For
this we bin the FIRST and NVSS catalogs in their entirety in logarithmic flux
bins to derive an estimate of the number of sources per flux interval and
solid angle, $\frac{dN}{dS \, d\Omega}$. Due to the large sample size of both
NVSS and FIRST, this quantity is well constrained except at very low and very
high flux densities. Because it is difficult to find a generally valid fitting
function, we model $\frac{dN}{dS \, d\Omega}$ using quadratic spline
interpolation between bins.

The number of foreground/background sources, 
%$N_{\mathrm{bg}}$, 
$N_{\eta}^{\mathrm{field}}$, inside the search radius 
%(defined as
%$\eta \, r_{200}$, where $\eta$ is a fraction of $r_{200}$) 
of a given cluster is determined by integrating over the \mtx{field
  density}\footnote{\mtx{We approximate the \textit{field} as the sum
    of all regions on the sky covered by the radio survey, minus
    circular regions around the sample clusters, defined by the
    respective radii ($r_{200}$) estimated from scaling relations.}}
as
\begin{equation}
%  N_{\mathrm{bg}} = \Omega  \int_{S_{\rm{min}}}^{S_{\rm{max}}} \frac{dN}{dS \, d\Omega} dS,
  N_{\eta}^{\mathrm{field}} = \Omega  \int_{S_{\rm{min}}}^{S_{\rm{max}}} \frac{dN}{dS \, d\Omega} dS,
\end{equation}
where $\Omega$ is the angular area of the searched region, and $S_{\rm{min}}$
and $S_{\rm{max}}$ are determined from the limiting values of the luminosity
bin according to Eq.~(\ref{eq:lumeq}). 
%We make the approximation that
%cluster sources do not significantly contribute to the mean surface number
%density.

The total number of cluster sources inside the search radius is
\begin{equation*}
N_{\eta}=N_{\eta}^{\rm{count}}-N_{\eta}^{\mathrm{field}}.
\end{equation*}
\mtx{Uncertainties on $N_{\eta}$ are computed from Poisson statistics
  as discussed by \cite{1986ApJ...303..336G}.}

For reasons of simplicity, the previous discussion has focused on source
counts in a single cluster field. However, it is straightforward to apply the
method discussed here to the entire stack of clusters by adding all individual
$N_{\eta}$.

\subsubsection{De-projection into cluster volume}
\label{sec:lf:method:deproj}

To convert the cluster source counts $N_{\eta}$ inside the ``line-of-sight
cylinder'' (cf. Fig.~\ref{fig:lf:cartoon}) limited by $\eta \, r_{200}$ to
counts $N$ inside $r_{200}$, we compute the expected ratio of sources inside
the sphere delimited by $r_{200}$ and inside the cylinder. We integrate the
spatial source density, as a function of de-projected radius, over the two
regions and take the ratio
\begin{equation}
  C_N(\eta)  = \frac{N_{\text{sphere}}}{N_{\text{cylinder}}} =
 \frac{4 \pi \int \limits_{\Xi=0}^{1} \Psi(\Xi) \, \Xi^2 \, d\Xi}
 {2 \pi \int \limits_{z=-\infty}^{+\infty} \int \limits_{\rho=0}^{\eta} 
 \Psi\left(\sqrt{\rho^2 + z^2} \, \right) \, \rho \, d\rho dz}, 
\label{eq:deproject}
\end{equation}
where $\Psi(\Xi)$ is the de-projected counterpart of $\psi(\xi)$
(cf. Eq.~\ref{eq:proffitfun}). Note that $\Xi = r/r_{200}$, with $r$
now being the de-projected (physical) coordinate. The integral over
the line-of-sight cylinder (in the denominator) is written in
cylindrical polar coordinates, with $\sqrt{\rho^2 + z^2} = \Xi$.

\mtx{The de-projected source distribution $\Psi(\Xi)$ is derived from
  the projected source density distribution $\psi(\xi)$, discussed in
  \sect\ref{sec:profiles}. Because no dependence on luminosity could
  be determined for $\psi(\xi)$, we use the parameters derived from
  the total sample for each combination of cluster and radio source
  catalogs, as listed in Table~\ref{tab:profres}.}

Because the central Gaussian component of $\psi(\xi)$ in
Eq.~(\ref{eq:proffitfun}) does not correspond to a physical
distribution of sources (cf. \sect\ref{sec:profiles}), it is not
physically meaningful to de-project it to physical
coordinates. However, because the component is narrow, we can make the
approximation that all sources belonging to it (i.e.  all BCGs) have
been counted inside the line-of-sight cylinder, provided $\eta$ is
chosen large enough ($\eta \gtrsim 0.1$).  Thus, no correction in the
sense of Eq.~(\ref{eq:deproject}) is required for the central
component.

\mtx{We compute a correction factor only for the broader
  $\beta$-model.  In order to avoid over-correction where the narrow
  component is present, we compute from (\ref{eq:proffitfun}) the
  relative fraction of sources belonging to the broader profile and
  multiply the correction factor by this number to yield a modified
  correction factor, $\tilde{C}_N$}.

For the isothermal $\beta$-model, the physical (de-projected) density
of the sources is given by \citep[e.g.][]{1988xrec.book.....S}
\begin{equation}
\Psi(\Xi) = \Psi_{\beta} \left( 1+ \frac{\Xi^2}{\zeta^2} \right) ^ {-\frac{3}{2}\beta},
\label{betavol}
\end{equation}
which is the function we use when computing the integrals in
Eq.~(\ref{eq:deproject}). The parameters $(\beta,\zeta)$ are unique
for each galaxy cluster sample, as discussed in
\sect\ref{sec:profiles}. Because we are only interested in ratios, we
set $\Psi_{\beta}=1$.

The cluster volume is defined as a sphere with radius $r_{200}$. Thus, the RLF
in luminosity bin $\Delta L$ can be expressed as
\begin{equation}
\phi(L) \Delta L \ = \tilde{C}_N \frac{N_{\eta}}{V_{\rm{sph}}},
\end{equation}  
where $V_{\rm{sph}}$ is the sum of physical cluster volumes (within
$r_{200}$), and $N_{\eta}$ is the total number of sources (in all
cluster fields) found within a radius $\eta \, r_{200}$ and having
luminosities in the luminosity bin $\Delta L$.

The uncertainty in the RLF is dominated by Poissonian noise in the
source counts. We compute the uncertainty by scaling from the
uncertainty $\delta N_{\eta}$ in $N_{\eta}$ as
\begin{equation*}
\frac{\delta (\phi(L) \Delta L)}{\phi(L) \Delta L} = 
\frac{\delta N_{\eta}}{N_{\eta}}.
\end{equation*}  

By construction, the RLF is insensitive to the chosen value of the projected
fractional radius $\eta$ after the correction factor $\tilde{C}_N$ has been applied.
However, uncertainties in the RLF can be minimized by carefully tuning this
parameter. Values too close to 1 will increase the background counts, and
thereby the error, since there are few cluster sources close to $r_{200}$. On
the other hand, too low values of $\eta$ will increase uncertainties due to
not providing enough cluster sources for good statistics. We find that
$\eta=0.5$ provides a good balance between cluster and background sources, and
use this value in computing the RLF.

\subsection{Confusion}
\label{sec:lf:confusion}

Confusion becomes important when the typical separation of sources is
comparable to the angular resolution of the radio survey from which the RLF is
derived. In particular, a larger beam tends to overestimate the RLF at the
high luminosity end as several confused low luminosity sources appear as fewer
sources with higher luminosity. This effect is redshift dependent through the
conversion of angular to physical distance, which complicates quantifying a
possible redshift evolution in the RLF.

Comparing the narrow radial distribution of sources (section
\ref{sec:profiles}) with the resolution of the FIRST survey, it is
apparent that confusion affects the counts also when using FIRST data.
This will affect the shape and normalization of the RLF in a way that
we have no direct way of quantifying.  However, our main interest is
to constrain the redshift evolution of the RLF, and for this purpose
the absolute normalization of the RLF is of less importance.
Therefore, rather than attempting to correct for confusion, \mtx{we
  can correct for the relative difference in confusion at different
  redshifts.}

\subsubsection{Degrading the resolution of the radio data}
\label{sec:lf:confusion:degrade}

To address the confusion problem, we thus degrade the resolution of the
radio source data in an adaptive way to make sure that confusion effects are
the same at all redshifts (disregarding second order effects such as a number
density evolution in the central parts of clusters influencing the confusion
problem).  In the following, we discuss how the resolution of the radio source
data can be degraded. To demonstrate the method, we apply it to the FIRST
catalog to enable a direct comparison of the RLF derived from FIRST to that
derived from NVSS, before discussing how the method is used in removing
first-order systematic effects of confusion in the redshift evolution of the
RLF.

The full-width half maxima (FWHM) of the FIRST and NVSS synthesized beams are
5$\arcsec$ and 45$\arcsec$, respectively. To what extent individual sources
can be distinguished in each of the surveys, however, largely depends on their
absolute and relative brightnesses. \mtx{Here we make the simplification
  that given a resolution in terms of a beam FWHM, no two sources are resolved
  if their mutual separation (irrespective of flux) is less than the FWHM.}

To degrade the resolution of the FIRST or NVSS data, we define a beam
with a FWHM larger than that of the original data. In a given field,
the position and flux density of each source are recorded. The
brightest source is located, and all sources within a radius equal to
the FWHM around this source are combined into one source. The position
of the new source is taken as a flux-weighted average of its parts,
and the total flux density is the sum of the integrated flux densities
of the parts. The next brightest source (excluding all sources that
have already been considered) is then located, and the procedure is
repeated until no two sources are separated by an angular distance
less than the FWHM. Note that this method is limited by the use of
source catalogs in the sense that sources fainter than a chosen limit
will not be considered. Here we use the completeness limit of the two
source catalogs to determine which sources to consider for summing.

As an improved method, one could instead use the raw maps of cluster
fields from the FIRST survey and degrade them to the NVSS resolution
before extracting sources. However, this requires a robust source
extraction in both maps and is quite complicated for faint sources
near the completeness limit of the surveys. \mtx{Moreover, neither
  such a method nor the method described above can resolve the
  inherent problem that some extended emission recovered by NVSS is
  resolved out by FIRST.}

\subsubsection{NVSS and FIRST source counts}
\label{sec:lf:confusion:nvssfirst}

A simple way to visualize the effect of confusion on the RLF is to
compare the luminosity function as constructed from NVSS with that
constructed from FIRST, using the same completeness cutoff in both
surveys. For this comparison we select cluster fields in a narrow
redshift range, $0.1 \leq z \leq 0.16$, from the maxBCG catalog to
ensure a sufficiently large statistical sample. \mtx{This
  sub-sample of the maxBCG catalog is used only for the purpose of
  directly comparing luminosity functions constructed from FIRST and
  NVSS in this section. It contains 2341 clusters with $M_{200} > 5
  \times 10^{13} M_{\odot}$ which are covered by NVSS and FIRST.}  As
indicated in Fig.~\ref{fig:nvss:first}, computing the RLF directly
from the two radio source samples yields inconsistent results. This is
expected due to confusion effects.

We attempt to re-create the NVSS based luminosity function from the
FIRST data by degrading the resolution as described above. \mtx{Note
  that we do not expect a perfect agreement since the sensitivity to
  spatial frequencies of FIRST is very different from that of NVSS. }

Because both surveys have many sources separated by distances much
smaller than the respective resolutions (due to the specific source
extraction methods applied to the synthesized images), we impose a
strict lower limit of 45$^{\prime\prime}$ on the separation of sources
in \textit{both} surveys for this comparison. The results are shown in
Fig.~\ref{fig:nvss:first}.
\begin{figure}[ht!]
  \centering \includegraphics[width=\columnwidth]{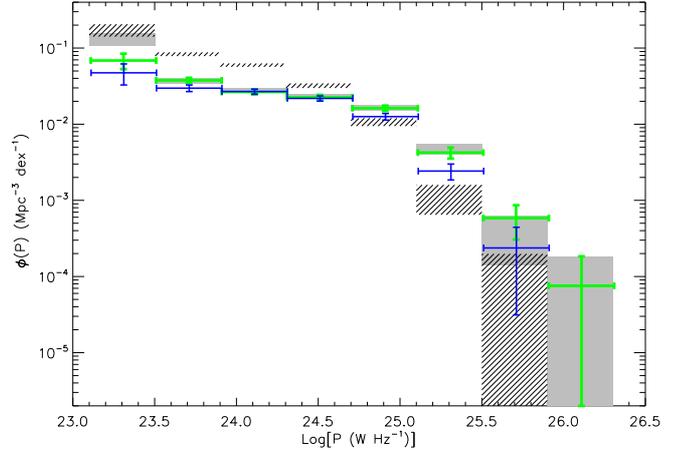}
  \caption{Change in the RLF resulting from degrading the resolution
    of both the FIRST and NVSS catalogs to a minimum source separation
    of 45$^{\prime\prime}$ (the FWHM of the synthesized beam of
    NVSS). The shaded regions indicate the RLF computed from the
    non-degraded NVSS (grey) and FIRST (hashed) data.  The error bars
    indicate the degraded versions; here the counts derived from the
    FIRST catalog (blue, thin error bars) are in approximate agreement
    with the NVSS counts (green, thick error bars), despite the large
    difference in synthesized beam FWHM of the original data.}
\label{fig:nvss:first}
\end{figure}
Even using this simple method, the RLF from the degraded FIRST sample
is in good agreement with that derived from the NVSS counts. The RLF
constructed from the degraded FIRST data is systematically lower than
that constructed from the degraded NVSS data. \mtx{This is as
  expected considering the different sensitivities to spatial
  frequencies of the two surveys $-$ extended emission recovered by
  NVSS is resolved out by FIRST, causing the amplitude of the RLF to
  drop.} Note, however, that the difference in the luminosity function
at different redshifts caused by this effect will in fact be much
smaller than the residual difference seen in
Fig.~\ref{fig:nvss:first}, \mtx{since our relatively small
  redshift range in the maxBCG sample results in a much smaller
  relative difference in physical scales, as compared to the relative
  difference between the FIRST and NVSS synthesized beams.}

\subsubsection{Accounting for confusion}
\label{sec:lf:confusion:accounting}

Given the above model, it is possible to adaptively introduce
confusion into our radio source sample to minimize systematic effects
in determining the redshift dependence of the RLF. In a given cluster
sample, the clusters with the greatest angular diameter distances will
be the most affected by confusion.  Thus, we define a nominal
``confusion distance'', in physical units, by converting the radio
source survey resolution (45$^{\prime\prime}$ for NVSS or
5$^{\prime\prime}$ for FIRST) to a physical distance at the redshift
of the object with the greatest angular diameter distance in the
sample. Then, for every object in the sample, this physical distance
is converted back to angular units using the redshift of the object,
and the resulting angular scale is used to degrade the resolution of
this particular cluster field using the method described above.

To get an idea of the typical level of confusion, consider the maximum
redshift, $z=0.3$, of the maxBCG catalog. The FIRST resolution of
5$^{\prime\prime}$ corresponds to a physical distance of 22 kpc at
this redshift. We can translate this physical distance, the
``confusion distance'', into an angular distance at any lower
redshift; for example, at the low redshift limit of $z=0.1$ we obtain
an angular scale of 12$^{\prime\prime}$. Using this as a limit of
resolution, the FIRST source counts are reduced by approximately 25\%
in fields defined by $r_{200}$, while at $z=0.3$ the FIRST source
counts are reduced by around 10\%.\footnote{The counts are reduced
  also at the resolution limit because, contrary to the FIRST catalog,
  we enforce a strict limit of 5$^{\prime\prime}$ as the smallest
  angular distance between two distinct sources.}

Note that we also have to adapt the background counts,
$N_{\eta}^{\mathrm{field}}$, when degrading the resolution.  Because it is
time consuming to re-compute $\frac{dN}{dS \, d\Omega}$ for each angular
resolution used, we compute the field counts in several degraded versions (as
described in \sect\ref{sec:lf:confusion:degrade}) of the entire NVSS and
FIRST catalogs, using increments of 5$^{\prime\prime}$, and interpolate
between these results to derive estimates of $N_{\eta}^{\mathrm{field}}$
individually for all cluster fields.

\subsection{Modeling the luminosity function}
\label{sec:lf:model}

\subsubsection{Parametric model}
\label{sec:lf:model:param}

To allow for a quantitative estimate of the redshift evolution of the
RLF, we follow LM07 and use the parameterization
\begin{equation}
  \log \phi = y - \left( b^2 + \left( \frac{\log L - x}{w} \right) ^2 \right) ^{1/2} -
  1.5 \log L.
\label{eq:condonmodel}
\end{equation}
This fitting function was used by \cite{2002AJ....124..675C} to fit
the field RLF using the combined contributions from radio-loud AGN and
star-forming (SF) galaxies.  LM07 used the two sets of values of
($b,x,w$) found by Condon et al. to make possible a separation of the
normalizations ($y$) of the radio-loud AGN and SF components in their
cluster RLF at low redshift.

\mtx{ It is beyond the scope of this paper to discuss the physical
  interpretation of Eq.~(\ref{eq:condonmodel}) as our main concern is
  the redshift evolution of the luminosity and number density. Refer
  to \cite{2002AJ....124..675C} and references therein for details on
  this hyperbolic fitting form of the RLF.}

\mtx{In contrast to the sample studied by LM07, our main samples do
  not include systems with $z < 0.1$. This selection, combined with
  our flux limit of 5 mJy, effectively ensures that we are only
  sensitive to the high-luminosity radio-loud AGN population of the
  RLF. Thus, a one-component model as described by
  Eq.~(\ref{eq:condonmodel}) is sufficient for the present purpose.}

\subsubsection{Redshift evolution}
\label{sec:lf:model:zevol}

\cite{2000A&A...360..463M} quantified the redshift evolution of the
RLF \mtx{under the assumption that the overall shape remains constant,
  as first suggested by \cite{1984ApJ...287..461C}. Under this
  assumption, there can only be changes in overall luminosity and
  overall number density}.  The redshift dependence can be written as
\begin{equation}
\phi(L, z) \ = \ g(z) \ \phi \left[Lf(z), z\approx 0 \right],
\label{eq:lf:zevol}
\end{equation} 
where $g(z)$ quantifies the number density evolution, corresponding to a
vertical shift $\Delta Y$ in $\phi(L)$, and $f(z)$ represents luminosity
evolution, corresponding to a horizontal shift $\Delta X$:
\begin{equation}
\log\left[ \phi(\log L,z)\right] \ = \ \log\left[\phi(\log L + \Delta X, z \approx 0)\right] \ + \ \Delta Y.
\label{eq:lf:zevol_shift}
\end{equation} 
The vertical and horizontal shifts can be fitted to yield $f(z)$ and $g(z)$.
Shape preservation of the RLF in ($\log L, \log \phi$) space implies that
number density and luminosity scale as powers of redshift. For the luminosity,
\begin{equation}
\label{eq:lf:ppowerlaw}
L=L_0\left( \frac{1+z}{1+z_0} \right)^{\alpha_L},
\end{equation}
where $\alpha_L$ is the power law index, and correspondingly for the number
density
\begin{equation}
\label{eq:lf:phipowerlaw}
\phi = \phi_0\left( \frac{1+z}{1+z_0} \right)^{\alpha_{\phi}},
\end{equation}
with the power law index $\alpha_{\phi}$.

To constrain the redshift evolution in the RLF, we bin our samples by
redshift, and approximate the redshift in each bin by the median of
all cluster redshifts in the bin. We then apply
(\ref{eq:condonmodel}), shifted in $\phi$ and $L$ according to
(\ref{eq:lf:zevol_shift}), to constrain the power law indices
$\alpha_{\phi}$ and $\alpha_L$. Simultaneously, the four parameters of
(\ref{eq:condonmodel}) are constrained.

We let the parameter $y$ in (\ref{eq:condonmodel}) represent the
normalization of the RLF at $z=0$. We bin the data by redshift and
construct the RLF separately for each redshift bin. The thus
constructed luminosity function data are then used simultaneously to
constrain $y,b,x,w,\alpha_{\phi}$ and $\alpha_L$.  \mtx{The fitting is
  carried out by minimization of the $\chi^2$ statistic,
\begin{equation}
  \chi^2 = \sum_{i=1}^{k}\left( \frac{X_i - \mu_i}{\sigma_i} \right) ^2,
\end{equation}
where $X_i$ is the data point with index $i$ (the number $k$ of data
points is the total number of luminosity bins in all redshift bins),
$\mu_i$ is the corresponding model value given a set of parameters
($y$, $b$, $x$, $w$, $\alpha_{\phi}$ and $\alpha_{L}$), and $\sigma_i$
is the uncertainty on data point $i$ estimated from Poisson
statistics.  }

\mtx{Uncertainties on the fitted parameters are estimated
  using a simple method. Given our data points, binned in redshift and
  in luminosity, we vary the data according to Poisson statistics
  within their estimated uncertainties, and carry out the $\chi^2$ fit
  again. This process is repeated 10.000 times. As expected, the mean
  of the distribution of a fitted parameter in any of the fits we
  carry out (\sect\ref{sec:lf:res:zevol}) corresponds to its best-fit
  value. The standard deviation of the distribution of a fitted
  parameter is used as an estimate of the 1$\sigma$ uncertainty of
  this parameter.}

\subsubsection{Mass dependence}
\label{sec:lf:model:mdep}

A possible mass dependence in the RLF can be investigated using the
same method as described above. By scaling the number density and
luminosity we can model the dependence to a first approximation as
power laws of the form $M_{200}^{\gamma}$. \mtx{Analogously with the
  discussion in \sect\ref{sec:lf:model:zevol}, we make the assumption
  of shape preservation of the RLF} and model the luminosity and
number density dependence on mass as (cf. Eqs.~(\ref{eq:lf:ppowerlaw})
and (\ref{eq:lf:phipowerlaw}))
\begin{equation}
\label{eq:mdeppower}
\begin{split}
L \sim (M_{200})^{\gamma_L};  \\
\phi \sim (M_{200})^{\gamma_{\phi}}.
\end{split}
\end{equation}

%As discussed in \sect\ref{sec:samples:clusters}, mass estimates for
%X-ray selected clusters are far more reliable than for the optical
%sample.  
We use both the X-ray sample and the optical sample to constrain the
mass dependence. Because two new parameters are introduced into the
model, it is not possible to simultaneously constrain the mass
dependence and the redshift evolution, particularly given the limited
size of the X-ray sample. For this reason we use a sub-sample of the
X-ray selected clusters in a low redshift range of $0.05 \leq z \leq
0.12 $ (see Table \ref{tab:ccuts}). In this range, we divide the
sample into two mass bins, where care is taken that the lower mass
limit of the lower mass bin is above the completeness limit of the
X-ray sample at $z=0.12$. For the optical sample, we divide the
complete sample into three mass bins. In the next section it is shown
that in this range, our data are consistent with no mass dependence of
the RLF.

\subsection{Results}
\label{sec:lf:res}

\subsubsection{Mass dependence}
\label{sec:lf:res:mdep}

To separate a possible mass dependence in the RLF from a pure redshift
dependence, we construct RLF binned by mass as shown in
Figures~\ref{fig:mdep.xray} and \ref{fig:mdep.opt}.
\begin{figure}[Ht!]
  \centering \includegraphics[width=\columnwidth]{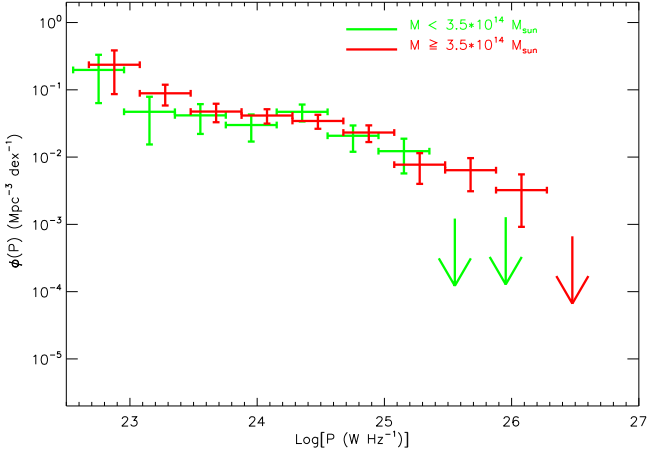}
  \caption{RLF determined from the X-ray subsample in the redshift
    range $0.05 \leq z \leq 0.12$, divided into two mass
    bins. \mtx{Arrows indicate $1 \sigma$ upper limits
      from Poisson statistics.} }
\label{fig:mdep.xray}
\end{figure}
\begin{figure}[Ht!]
  \centering \includegraphics[width=\columnwidth]{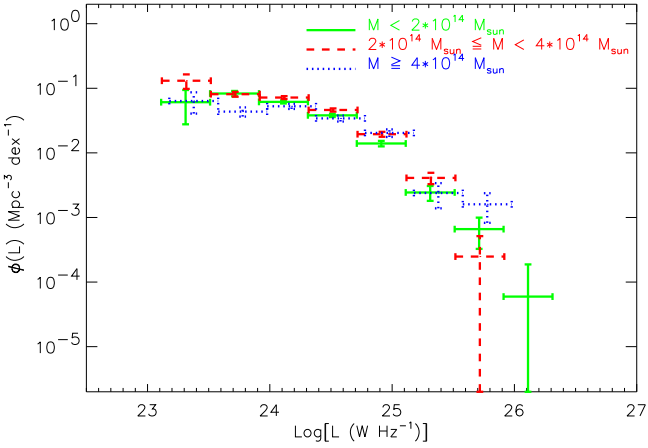}
\caption{RLF determined from the optical sample, divided into
  three mass bins.}
\label{fig:mdep.opt}
\end{figure}
We then use a least-squares statistic to find the best-fit vertical and
horizontal shifts of the RLF, fixing the values of $b,w,x$ as described in
\sect\ref{sec:lf:model:zevol}, to interpret the shifts in terms of a
luminosity- and/or number density dependence on cluster mass.  The results for
both our cluster catalogs are given in Table \ref{tab:mdepfits}, using
different priors on the power laws.
\begin{table}
\caption{Mass dependence of the RLF, parameterized in terms of power laws of
   $M_{200}$ in luminosity and number density (Eq.~(\ref{eq:mdeppower})).}
\label{tab:mdepfits}      
\centering                          
\begin{tabular}{l l r r@{}l r@{}l}
\hline\hline                 
Cluster & Source  & Prior             &    $\gamma_{\phi}$&  & $\gamma_{L}$&    \\  
sample  & catalog &                   &                   &  &             &    \\
\hline
X-ray   & FIRST   &                   &      0.25$\pm$&1.01  & -0.46$\pm$&0.55   \\ 
maxBCG  & FIRST   &                   &     -0.36$\pm$&0.31  &  0.38$\pm$&0.65   \\ 
\hline
X-ray   & FIRST   & $\gamma_{\phi}=0$ &       (0.0)~\,&      &  0.13$\pm$&0.36   \\ 
maxBCG  & FIRST   & $\gamma_{\phi}=0$ &       (0.0)~\,&      &  0.27$\pm$&0.28   \\ 
\hline
X-ray   & FIRST   & $\gamma_{L}=0$    &     0.090$\pm$&0.16  &   (0.0)~\,&       \\ 
maxBCG  & FIRST   & $\gamma_{L}=0$    &     0.046$\pm$&0.15  &   (0.0)~\,&       \\ 
\hline                                   
\end{tabular}
\end{table}
Note that all fits are consistent with $\gamma_{\phi}=0$ and $\gamma_{L}=0$ at
an approximate 1 $\sigma$ level.

Thus, contrary to LM07, who find that the amplitude of the RLF for
low-mass clusters is slightly larger at the low-luminosity end, we
find no evidence of a mass dependence in the RLF given our flux and
redshift cut-offs. As indicated in Figure \ref{fig:mdep.xray}, the
results are consistent with the most luminous radio sources only being
found in high-mass clusters, as was also found by LM07. However,
\mtx{due to the relatively small sample size, our data provides no
  conclusive evidence of this.}

As there is no statistically significant indication of a mass
dependence from either cluster sample, we proceed to fit for a pure
redshift evolution in the next subsection.

\subsubsection{Redshift evolution}
\label{sec:lf:res:zevol}

Figure \ref{fig:zevol.opt} 
\begin{figure}[Ht!]
  \centering \includegraphics[width=\columnwidth]{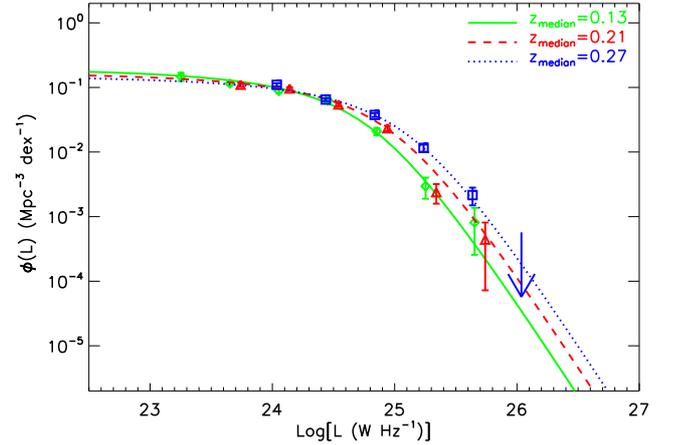}
  \caption{Redshift evolution in the RLF determined from the optical
    (maxBCG) sample, \mtx{without priors}. The data are shown as error
    bars and the best simultaneous fit to luminosity and number count
    evolution is represented by lines corresponding to the redshift
    bins: $0.1 \leq z < 0.17$ (diamonds and solid line), $0.17 \leq z
    < 0.24$ (triangles and dashed line), and $0.24 \leq z \leq 0.3$
    (squares and dotted line).  }
\label{fig:zevol.opt}
\end{figure}
shows the RLF computed from the maxBCG sample, using FIRST radio
sources, in three redshift bins chosen such that there are
approximately the same numbers of clusters in each bin ($0.1 \leq z <
0.17$, $0.17 \leq z < 0.24$ and $0.24 \leq z \leq 0.3$).  We apply the
method outlined in \sect\ref{sec:lf:model:zevol} to constrain the
redshift evolution of the RLF by a simultaneous fit to the RLF
amplitude (parameterized by $y$ at $z=0$), the RLF shape
(parameterized by $b$, $x$ and $w$), and the power law evolution
(parameterized by $\alpha_{\phi}$ and $\alpha_{L}$ according to
Eqs.~(\ref{eq:lf:ppowerlaw}) and (\ref{eq:lf:phipowerlaw})).

We perform a number of fits to the data, with priors as listed in
Table~\ref{tab:zevol}.
\begin{table*}
\begin{center}
\caption{Redshift evolution in the RLF, quantified by simultaneous fits to the
  RLF amplitude, shape and evolution. The goodness-of-fit is indicated by the
  reduced chi-squared parameter, $\chi_{\text{red}}^2$. See section
  \ref{sec:lf:model} for a description of the parameters.}
\label{tab:zevol}      
\centering 
\begin{tabular}{l l r r@{}l r@{}l r@{}l r@{}l r@{}l r@{}l r@{}l}
\hline\hline                 
Cluster & Source  & Priors            & $y$&            & $b$&           & $x$&            & $w$&           & $\alpha_{\phi}$&  & $\alpha_{L}$&   & $\chi_{\text{red}}^2$& \\
sample  & catalog &                   &                 &                &                 &                &                   &                 &          \\
\hline
maxBCG   & FIRST  &                   & 36.38$\pm$&1.02 & 1.05$\pm$&0.73 & 24.53$\pm$&0.18 & 0.66$\pm$&0.13 & $-$2.46$\pm$&1.58 & 6.20$\pm$&1.76  & 1.&07    \\
maxBCG   & FIRST  & $\alpha_{\phi}=0$ & 36.34$\pm$&0.92 & 0.91$\pm$&0.81 & 24.87$\pm$&0.14 & 0.72$\pm$&0.21 &    (0.0)~\,&      & 3.99$\pm$&1.24  & 1.&19     \\
maxBCG   & FIRST  & $\alpha_{L}=0$    & 36.74$\pm$&0.89 & 1.01$\pm$&0.55 & 25.11$\pm$&0.11 & 0.71$\pm$&0.19 &    1.03$\pm$&1.14 & (0.0)~\,&       & 2.&25    \\
\hline
X-ray    & FIRST  & (a)               & 36.19$\pm$&0.19 & (1.05)~\,&     & (24.53)~\,&     & (0.66)~\,&     &    0.76$\pm$&1.86 & 8.12$\pm$&2.67  & 0.&94    \\
X-ray    & FIRST  & (a);~$\alpha_{\phi}=0$ & 36.26$\pm$&0.10 & (1.05)~\,& & (24.53)~\,&    & (0.66)~\,&     &    (0.0)~\,&      & 8.19$\pm$&2.66  & 0.&89     \\
X-ray    & FIRST  & (a);~$\alpha_{L}=0$ & 35.89$\pm$&0.18 & (1.05)~\,& & (24.53)~\,&      & (0.66)~\,&     &    9.40$\pm$&1.85 & (0.0)~\,&       & 10.&48    \\
\hline                                   
\end{tabular}
\end{center}
(a) The shape parameters $b$, $x$ and $w$ were fixed to the results of the maxBCG/FIRST
analysis with no priors.  \\  
\end{table*}
Although a simultaneous fit to both $\alpha_{\phi}$ and $\alpha_{L}$
yields the best fit (in the sense of the reduced $\chi^2$ parameter,
$\chi_{\text{red}}^2$, being the closest to unity), the data are
consistent with no number density evolution.  Note that the best-fit
value $\alpha_{\phi}=-1.38$ implies \textit{fewer} radio-loud AGN at
higher redshift within the range $0.1 \leq z \leq 0.3$.  \mtx{Though
  not unphysical, such a negative evolution is unlikely, considering
  the evidence of increased AGN activity in the field population
  \citep{1990MNRAS.247...19D}, and the enhanced AGN fraction within
  clusters as seen from X-ray observations
  \citep{2007ApJ...664L...9E,2009ApJ...694.1309G}} \mtx{Therefore, we
  carry out an additional fit in which we fix the number density power
  law to zero} and fit for a pure luminosity evolution.  \mtx{The
  result is a positive luminosity evolution with $\alpha_L=3.99 \pm
  1.24$}.  While consistent with a pure luminosity evolution, the data
are inconsistent with a pure number density evolution, as indicated by
the $\chi_{\text{red}}^2$ of the third column in
Table~\ref{tab:zevol}.

The last three rows of Table~\ref{tab:zevol} indicate the results of
fitting the evolution parameters to the X-ray derived RLF, also binned
by redshift (redshift bins: $0.1 \leq z < 0.2$, $0.2 \leq z < 0.45$
and $z \geq 0.45$).  Although the X-ray sample extends to greater
redshifts than the maxBCG sample, it has less leverage on the redshift
evolution because of its limited size. For this reason,
\mtx{we fit only for $y$, $\alpha_{\phi}$ and
  $\alpha_{L}$.  We keep the shape parameters fixed to the values
  found in the maxBCG/FIRST fit, as this is the fit with the smallest
  uncertainties. Because the smallest $\chi^2_{\text{red}}$ is
  measured for the case of no priors (as expected) in the maxBCG/FIRST
  case, we use the results of this fit for fixing the shape
  parameters.}  Again, we also carry out fits to pure luminosity
evolution and pure number density evolution. At one standard
deviation, the results from the X-ray sample are consistent with those
of the optical maxBCG sample. Again, a pure number density evolution
provides a poor fit to the data.

We note that in all cases the fit is especially sensitive to the
parameter $b$, which is explained by its quadratic dependence in
Eq.~(\ref{eq:condonmodel}).

\subsection{Systematic uncertainties}
\label{sec:lf:sys}

\mtx{ To estimate systematic uncertainties in the luminosity
  and number density evolution of the RLF, we discuss below four
  separate sources of error. 

  \mtx{The estimated systematic uncertainties for the maxBCG and X-ray
    samples are summarized in Table \ref{tab:sys}. For each
    uncertainty estimate, we keep all parameters but the RLF
    normalization and the evolution parameter in question
    ($\alpha_{\phi}$ or $\alpha_{L}$) fixed, while carrying out the
    complete RLF analysis with posterior assumptions altered as
    described above. To estimate total systematic effects on
    $\alpha_{\phi}$ and $\alpha_{L}$, we add all relevant components
    in quadrature. } \mtx{We note that for this study, the statistical
    errors on the fitted parameters (see Table~\ref{tab:zevol}) still
    dominate over the systematics.}

\begin{table}[Ht]
\begin{minipage}[Ht]{\columnwidth}
\caption{Summary of systematic effects on the number density and
  luminosity evolution of the RLF.}             % title of Table
\label{tab:sys}      % is used to refer this table in the text
\centering                          % used for centering table
\renewcommand{\footnoterule}{}
\begin{tabular}{l l l}        % centered columns (4 columns)
\hline\hline                 % inserts double horizontal lines
Systematic & effect on $\alpha_{\phi}$ & effect on $\alpha_{L}$ \\    % table heading
\hline
$L_{\text{bol}}-T$ scaling & $\pm0.30$ & \\
$L_{X}-n_{\text{gal}}^{R200}$ scaling\footnote{only for maxBCG clusters} & $\pm$0.80 & \\
$k$-correction & $^{+0.15}_{-0.51}$ & $^{+0.17}_{-0.16}$ \\
\hline  
\hline                                 %inserts single line
Totals\footnote{added in quadrature} & & \\
\hline
maxBCG & $^{+0.87}_{-1.00}$ & $^{+0.17}_{-0.16}$ \\
X-ray & $^{+0.33}_{-0.59}$ & $^{+0.17}_{-0.16}$ \\
\hline
\end{tabular}
\end{minipage}
\end{table}

\subsubsection{Scaling relations}

Uncertainties in the $z$-dependence of the scaling relations used to
derive cluster masses affects the derived number density evolution
through the indirect determination of cluster volumes. For the X-ray
sample, we estimate this systematic by replacing the self-similar
evolution of the $L_{\text{bol}}-T$ relation by (i) the strong
evolution measured by \cite{2005ApJ...633..781K} and (ii) by assuming
no evolution of the $L_{\text{bol}}-T$ relation. Re-computing the
$M_{200}$ for both cases results in a change in the number density
evolution parameter $\alpha_{\phi}$ of $\pm0.30$ when re-computing the
RLF for the X-ray sample. For the optical sample, we additionally need
to take the uncertainty in the $L_{X}-n_{\text{gal}}^{R200}$ relation
(Eq.~(\ref{eq:rykoff})) into account. We estimate the systematic
effects by using the extreme values (at 1$\sigma$ confidence) of the
redshift dependence from \cite{2008ApJ...675.1106R} to re-compute
$M_{200}$, resulting in an effect on $\alpha_{\phi}$ of $\pm 0.80$.

\subsubsection{$k$-correction}

Uncertainties in the $k$-correction used in converting from radio
source flux to luminosity will mainly affect the luminosity
evolution. It can also shift the number density evolution by assigning
the wrong luminosity bins to sources. 

\mtx{As extreme values of the ensemble average spectral
  slope, we use the 25th and 75th percentiles ($\alpha = 0.51$ and
  $\alpha=0.92$, respectively) of the slope distribution of
  \cite{2007AJ....134..897C}.}  We re-compute the maxBCG RLF using
these extreme spectral slopes for all sources.

\subsubsection{Source counts}

Incorrect radio source counts can be caused either by counting
extended or complex sources (e.g. AGN with radio lobes) as several
separate sources, or by not resolving unrelated sources. This affects
the number density and luminosity evolution only if the effect is
varying with redshift; otherwise a mere offset in the normalization of
the RLF is introduced.

A complete treatment of these effects would require the careful visual
inspection of the radio maps in each cluster field, which is not
feasible for a large sample such as maxBCG. Instead we carried out a
visual inspection on parts of the sample to estimate the possible
effects.  We visually inspected FIRST maps in the 500 fields
corresponding to the clusters with the lowest redshifts, and in the
500 fields corresponding to the clusters with the highest redshifts of
the maxBCG sample, taking weighted sums of the flux densities of FIRST
catalog entries deemed to be different components or lobes of the same
sources. Ambiguous cases (less than 2\% of the fields studied) were
removed. Comparing the RLFs constructed from the modified FIRST data
in the two sub-samples, we noted a drop in normalization of the RLF by
about 5\% in both the high-$z$ and the low-$z$ subsamples. There was
no indication of a significant change to the evolution of the RLF
within our statistical errors. Thus, we conclude that our adaptive
accounting of confusion with redshift (section \ref{sec:lf:confusion})
is sufficient to deal with this systematic, at least for the purpose
of constraining the redshift evolution of the RLF.

\subsubsection{Radio source flux}

Systematic offsets in radio source flux densities are expected to be a
problem mainly for the FIRST survey, where significant amounts of flux
of extended sources can be resolved out. Again, this problem only
affects the normalization of the RLF, unless there is a redshift
dependence in the amount of resolved-out flux. Since the results from
the NVSS and FIRST radio data for the redshift evolution of the RLF
from the maxBCG sample are consistent within statistical errors, we do
not pursue this point further.

\subsection{Comparison of the RLF with previous findings}

\mtx{To compare our findings with previously published results, we
  consider the maxBCG/FIRST data in our lowest redshift bin, which
  compares well to the low-redshift samples of clusters used in
  previous works. Figure \ref{fig:lf.comparison}
\begin{figure}[Ht!]
  \centering \includegraphics[width=\columnwidth]{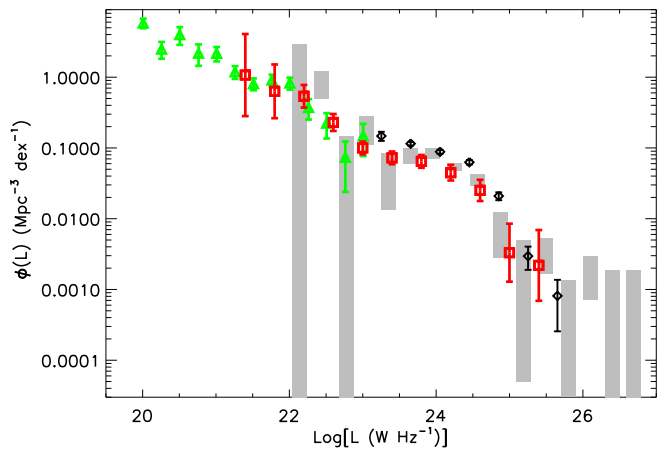}
  \caption{\mtx{Comparison of our RLF derived from the
      maxBCG/FIRST data in the lowest redshift bin ($0.1 < z < 0.17$,
      $z_{\text{median}}=0.13$; black diamonds) with other relevant
      volume-averaged radio luminosity functions. The RLF of
      \cite{2007ApJS..170...71L} is indicated by shaded regions, while
      the corresponding luminosity functions of
      \cite{2004ApJ...600..695R} and \cite{2004A&A...424..409M} are
      shown as green triangles and red squares, respectively.}}
\label{fig:lf.comparison}
\end{figure}
shows a comparison of our low-redshift RLF to the results of
\cite{2007ApJS..170...71L}, \cite{2004ApJ...600..695R} and
\cite{2004A&A...424..409M}.}

We note that the normalization of the RLF is consistent with the
results of LM07, \mtx{although the RLF from our optically selected
  sample does not extend as high in luminosity ($\textrm{Log}(L)
  \gtrsim 26 \, \textrm{W} \, \textrm{Hz}^{-1}$).  The latter} is
likely a result of the fact that the optically selected sample
contains many more low-mass clusters, and that the more luminous radio
sources tend to reside in high-mass clusters (cf.
\sect\ref{sec:lf:res:mdep}).

\mtx{The RLF of \cite{2004A&A...424..409M} has a
  significantly lower normalization than what we find,
  \mtx{by approximately a factor of two}. This can be
  expected due to the different definition of cluster volumes (using a
  constant cluster radius across the complete sample) used in the
  former work.}

\mtx{\cite{2006A&A...446...97B} used a definition of the
  RLF as the number of radio sources per cluster rather than averaged
  over a volume. A direct comparison with that work would require a
  re-analysis of the \cite{2006A&A...446...97B} sample, which would be
  only of moderate interest considering that the normalization would
  rely on different assumptions on cluster volume and the sample is
  selected differently from our maxBCG sample. Our main objective here
  is to show the approximate agreement between different local
  luminosity functions and to stress the difficulty in obtaining a
  robust normalization of the RLF; although the normalization depends
  strongly on the choice of cluster volume and is prone to systematic
  effects due to a poor understanding of this volume, this does not
  strongly affect our main conclusion, which is concerned with the
  redshift evolution of the RLF independently of its overall
  normalization.}

Early studies of the luminosity function of optically selected QSOs at $z \leq
2.2$ suggested a pure luminosity evolution with $L \sim (1+z)^{(3.5 \pm 0.3)}$
in the field population \citep{1987MNRAS.227..717B,1988MNRAS.235..935B}. A
later study of extragalactic radio sources by \cite{1990MNRAS.247...19D} at
2.7 GHz came to similar conclusions. 

The pure luminosity evolution of the volume averaged RLF found in this study
is approximately consistent with more recent findings in the field population
of radio galaxies at low and intermediate redshift \citep{2000A&A...360..463M,
  2001AJ....121.2381B}, although our best-fit evolutionary model with $L \sim
(1+z)^{(6.20 \pm 1.76)}$ is steeper than both the $L \sim (1+z)^{(4 \pm 1)}$
found by \cite{2001AJ....121.2381B} and the $L \sim (1+z)^{(3 \pm 1)}$ which
\cite{2000A&A...360..463M} found to be consistent with their data.

%%%%%%%%%%%%%%%%%%%%%%

\section{Summary and conclusions}
\label{sec:concl}

We used the maxBCG optical sample of galaxy clusters and a composite
X-ray sample to construct the volume averaged radio luminosity
function (RLF) in galaxy clusters by cross-correlating cluster
positions with radio point source positions from the FIRST and NVSS
survey catalogs. Background and foreground counts were corrected for,
and variable confusion with redshift was accounted for. We
investigated the radial source density distribution of radio sources
associated with clusters, \mtx{and correlated the luminosity of the
  brightest radio source with cluster mass}.

To fit the radial number density distributions, we used the
\mtx{functional form of the} isothermal $\beta$-model.  \mtx{Combining
  the maxBCG cluster sample with the FIRST catalog, we found an
  additional narrow component which we identify with radio detections
  of the brightest cluster galaxies (BCGs)}.  All combinations of
cluster sample (maxBCG or X-ray) and radio source sample (NVSS or
FIRST) yield similar values of the power law index $\beta$. The core
radius in the maxBCG/FIRST radial distribution of sources compares
well to the distribution of sources in the X-ray sample.  As it was
not possible to identify the narrow BCG component from the NVSS
survey, we found a much smaller core radius in the maxBCG/NVSS radial
distribution.

Our derived radial distributions are narrower than what has been found
in earlier studies \citep[e.g][]{2004ApJ...600..695R,
  2004A&A...424..409M}, \mtx{even when disregarding the
  central component associated with brightest cluster galaxies
  (BCGs)}. A plausible explanation is that we are constructing the RLF
at higher redshifts than previous studies and thus, given our flux
limit of 5 mJy, we are sensitive only to radio-loud AGN.  At lower
redshift there is a mixing with star forming galaxies, which have a
less \mtx{centralized} distribution.

\mtx{Unlike LM07, we do not find any evidence that bright radio sources
  have a radial source density distribution different from that of faint sources.
  Again, a likely explanation is that we are studying different populations of
  radio sources through the redshift selection.}

We found that the luminosity of the most radio-luminous source within 50 kpc
from the cluster center scales with cluster mass following a power law with
slope 0.31$\pm$0.12 in the maxBCG sample. This is consistent with the results
of \cite{2004ApJ...617..879L} as well as with the results from our X-ray sample,
although the latter is also consistent with no correlation.

\mtx{We find the RLFs constructed from the optical and X-ray samples of
  galaxy clusters to be in approximate agreement. The RLF from the optical
  maxBCG sample is systematically lower at luminosities $L \gtrsim 3 \times
  10^{25}$ W Hz$^{-1}$. This is likely a result of many more low-mass systems
  being present in the optical sample. }

We provide the first evidence for a luminosity evolution of the
volume-averaged RLF in clusters of galaxies.  The maxBCG/FIRST data
are consistent with a pure luminosity evolution, with power scaling
with redshift as $L \sim (1+z)^{\alpha_L}$, where $\alpha_L=6.20
^{+1.76 +0.19}_{-1.76 - 0.17}$ (statistical followed by systematic
uncertainties).  There is no indication of a mass dependence in the
RLF from the present data. However, the results from the X-ray sample
are consistent with the findings of LM07, that the most luminous radio
sources reside in massive clusters. This is further corroborated by
the fact that the RLF constructed from the maxBCG sample (which
contains a smaller fraction of high-mass systems than both our X-ray
sample and the sample of LM07) is steeper at higher luminosities.

\mtx{Below $P \sim 10^{25} \, \text{W} \, \text{Hz}^{-1}$, the data are consistent with no mass
dependence in the RLF, as shown by constructing the RLF in the
low-redshift X-ray sample and binning by mass. Although we have found
that massive clusters have more luminous BCGs, this effect is
counteracted in the RLF by the fact that these massive systems also
have more volume.}
Both the X-ray sample and the maxBCG sample yield results
consistent with a pure luminosity evolution of the RLF. In addition,
the derived power laws are comparable, although the samples cover very
different redshift ranges.

\begin{acknowledgements}
  We acknowledge partial support for this work from Priority
  Programme 1177 and Transregio Programme TR33 of the German Research
  Foundation (Deutsche Forschungsgemeinschaft).  For the early stages of this
  work, MWS acknowledges support through a stipend from the International Max
  Planck Research School (IMPRS) for Radio and Infrared Astronomy at the
  Universities of Bonn and Cologne.  FP acknowledges support from grant
  50\,OR\,1003 of the Deutsches Zentrum f\"ur Luft- und Raumfahrt. HA is
  grateful to Mexican CONACyT for research grants 50921-F and 118295.
\end{acknowledgements}

%%%%%%%%%%%%%%%%%%%%%%

\bibliographystyle{aa}   %% bibtex
\bibliography{martin}{}    %% bibtex

\end{document}